# Improving Prostate Gland Segmentation Using Transformer based Architectures.

Shatha Abudalou[1,2], Jung Choi[3], Yasin Yilmaz[2], Yoganand Balagurunathan[1,2*].

[1]Department of Machine Learning, [3]Diagnostic Radiology, H. Lee Moffitt Cancer Center and Research Institute, Tampa, Florida, USA.

Email:SA(shathaa@usf.edu),JC(jung.choi@mofitt.org),
YY(yasiny@usf.edu), YB (yoganand.balagurunathan@moffitt.org)

**Abstract:** Inter-reader variability and cross-site-domain shift are the major challenges in automatically segmenting prostate anatomy utilizing T2-weighted MRI images. We investigated this study to see if transformer models can retain their precision while accommodating such heterogeneity. Based on 546 MRI (T2-weighted) volumes annotated by two independent experts, we aim to compare the performance of the UNETR and SwinUNETR in prostate gland segmentation with our previously reported 3D U-Net model [1]. Three training strategies were studied, including the single cohort training dataset, 5-fold cross-validated mixed training cohort, and gland-size based training dataset. The hyperparameters were tuned by Optuna. The test set came from an independent population of readers and was used as the endpoint for the evaluation (Dice Similarity Coefficient). In our experiments on single reader training, SwinUNETR achieved an average dice score of 0.816 for Reader#1 and 0.860 for Reader#2. While UNETR achieved an average dice score of 0.8 and 0.833 for Reader#1 and Reader#2, respectively, compared to our baseline UNet with its average dice at 0.825 for Reader #1 and 0.851 for Reader #2. SwinUNETR features an average dice score of 0.8583 for Reader#1 and 0.867 for Reader#2 using cross-validated mixed training.

For the gland size based training dataset, SwinUNETR achieved an average dice of 0.902 for the Reader#1 subset and 0.894 for the Reader#2 subset by using five-fold mixed training strategy of (Reader#1, n=53 and Reader 2, n=87) at a large gland size-based subsets, where UNETR model showed a considerably poorer performance. We demonstrate that these global and shifted-window self-attention mitigate the label noise and class imbalance sensitivity resulting in average dice score improvements over CNNs of up to five points whilst maintaining computational efficiency. This contributes to the high robustness of clinical deployment of SwinUNETR.

## 1. Introduction

Prostate cancer is the second most common cancer in the US and worldwide and a large number of deaths from this type of tumor occur worldwide. The prostate gland must be delineated using Magnetic Resonance Imaging MRI in order that prostate cancer can be detected early. Delineating the prostate gland tumor in the first stages facilitates biopsies and helps the physicians to provide a treatment plan for treatment process and monitoring the patient's situation. However, the manual contouring is labor-intensive and time-consuming, it is also affected by the reader's delineation variability. Abudalou et al.[1] showed that a 3D UNet trained on data from human readers gives the average dice score in the range between 0.82 and 0.85. There is a shortcoming when applying these algorithms, especially with small real-world datasets, as this type of model cannot be generalized to all cases.

Convolutional Neural Networks (CNNs), have local connectivity and share weights make these networks more efficient in preprocessing the high dimensional medical imaging [2], adaptive to different segmentation tasks[3] and show great performance in local texture recognition; however, it is challenging to capture the context of longer range when the prostate accounts for less than 3% of an image's volume [4] . This problem is avoided by Vision Transformers (ViTs) via global self-attention. Models like TransUNet [5] combines the two-dimensional bottleneck from a ViT with a U-Net architecture, which increases range but overlooks in-plane information and does not scale well to an entire three-dimensional volume. UNETR [6] substitutes the three dimensional encoder with a purely ViT transformer-based architecture, however, its quadratic memory growth imposes restrictions on the native-resolution input. SwinUNETR [7] proposes shifted-window attention to reduce memory imbalance. Other variations like nnFormer[8] and TransBTS [9] compromise certain accuracy for increased efficiency by using axial or channel-wise attention but do not have multi-reader evaluation. Additionally, recent hybrid CNN-Transformer models have started to surpass the performance of pure CNN baselines across a variety of medical image segmentation benchmarks [10-14].

Where in [15] the authors applied the ViT into 3D U-Net model to the embedding of the U-Net architecture to refine the features without the need to down sampling the input images. In [16], the study presented the DT-VNet merging the transformer with VNet frame work and evaluated on three public prostate dataset, where the mean dice score for the Promise12 dataset 91.89.

Numerous studies have investigated the inter-reader variability for prostate gland segmentation. In [17] the authors utilized three deep learning models to segment the prostate gland with annotations by four readers reported dice similarity score from 0.84 up to 0.91. The study in [18], the authors employed the UNet model and five separate cohorts of expert annotations for studying the effect of label error. Model accuracy was determined to be such that a consensus of several predictions achieved greater than the human consensus. In[19], DenseNet and U-Net segmented the prostate gland using a model trained on 141 subjects and tested in 47 with a dice coefficient of 0.92. Where [20] U-resNet was used to segment glandular structures for two readers to compare the resulted model performances to obtain the best dice coefficient of 0.88 for central gland. Drawbacks of this work are based on training on a single cohort and small test data, which may impact generalization.

Despite advanced processes for the transformer-based models, they were trained on a single cohort dataset and evaluated on unseen datasets. Furthermore, the literature documents advancements for optimized prostate gland segmentation having been reported in the literature, various challenges remain unsolved, including class imbalance due to small size gland volume, the stability of models when only one MRI modality (T2-w) is provided and reducing variability in inter-reader datasets. To address these limitations and extending the findings of limited inter-reader generalizability of a 3D U-Net baseline model for the whole prostate gland segmentation, which has recently been demonstrated this work conducted to employ Transformer-based encoders, tailored and applied for medical image segmentation using effective 3D self-attention and domain adaptation schemes, where we thoroughly applied multiple training strategies to reach our generalization. We now provide the first work to perform a systematic benchmark of Transformer-based 3D architectures, specifically UNETR and SwinUNETR, all on the same multi-reader MRI dataset. Here, we perform the first in-depth analysis of UNETR and SwinUNETR on a 546-volume, inter-reader T2-weighted MRI dataset. Three experimental setups are implemented to demonstrate the robustness or limitations of existing approaches: a) single-cohort training to evaluate sensitivity to an individual "annotator style"; b) mixed-cohort cross-reader 5-fold cross-validation to assess cross-reader generalization; and c) gland-size stratification for the study of long-distance context capture. Optimization of the model hyperparameters is done using the Optuna framework, where the Dice similarity coefficients of the reader-specific sets constitute the endpoint of our evaluation.

In this work, we examine the performance of a state-of-the-art Transformer-based encoder in reducing variability in inter-reader datasets and adapting to multiple training strategies. In addition, we compared the performance of the applied models with the previous 3D UNet model performance, and we studied the strength of each model through the application of statistical analysis to determine if the attention mechanism can outperform the convolutional neural network for the prostate gland segmentation, with a focus on class imbalance, and the performance stability based on the gland size-based variability problems.

## 2. Materials and Methods:

In this section, we describe the pipeline to use transformer models with T2w MRI of two different readers to train, and evaluate the performance of trained UNETR and SwinUNETR for the prostate gland segmentation:

**Applied Model architectures** – Introduce the main differences of the compared model's architecture the UNet, UNETR, and SwinUNETR, along with a detailed description of the encoders and their decoder blocks.

**Datasets** – Describing the imaging sources, voxel spacing, and inert-reader class labels for the 546-volume cohort.

**Pre-processing and data augmentation** – resampling to isotropic resolution, intensity normalization, random flipping, cropping, and 3-D patch extraction.

**Loss functions and optimization** – The loss function is a mixture of a weighted Dice and cross-entropy using the weights to be the inverse class frequency. Learning-rate, batch-size, and feature-size hyperparameter tuning is conducted using Optuna with limited and predefined values of each of them.

**Evaluation protocol** – The evaluation protocol of training, and testing sets for the single-reader, the mixed-reader experiments, and the gland size-based dataset, with applying the primary evaluation metrics in the medical imaging segmentation the Dice similarity coefficient for the prostate gland segmentation. In addition, we used p-values among the applied models to make descriptive statistics comparisons between the performance distributions of these models and the experimental training strategy.

### 2.1 Model architecture:

In this study, we evaluated the performances of two Transformer-based encoders, 3D segmentation networks, SwinUNETR and UNETR, that employed different attention mechanism, and compared their performance to the baseline 3D U-Net in the previous work [1]. All these three models adopt an encoder–decoder architecture with skip connections but they differ essentially in their ways of getting contextual features. The encoder in the 3D U-Net model consists of a series of stacked convolution-pooling blocks. This architecture grew as a function of the number of layers used; here, the corresponding field also grew linearly with depth. The indirect inferences are needed only for the long-range dependencies [21]. However, the UNETR replaces such an encoder with a standard Vision Transformer, and the input volume was partitioned into non-overlapping $16^3$-voxel patches, which were then flattened and linearly embedded, before feeding into a stack of multi-head self-attention layers. Four layers of token-level embeddings extracted from different stages in the transformer are reshaped into 2D or 3D feature map and integrated with the decoder's skip paths this allow the transformer to be exposed to high resolution details from early stages ; thus the transformer decoder access to multi-scale global context at different resolutions [6]. SwinUNETR keeps the same decoder architecture but replaces the fully self-attentive encoder with a shifted-window Swin Transformer hierarchy. The self-attention is calculated in local $7^3$ windows that slide across stages where the input image divided into smaller non-overlapping patches and apply self-attention within each window not globally, providing almost linear memory growth while effectively allowing shift-window mechanism allow each new window spans across boundaries, then the model can efficiently build a global context layer-by-layer [7]. Accordingly, the Swin encoder used in the SwinUNETR model is significantly more memory- and computationally efficient than that of the UNETR quadratic-cost ViT encoder, thus the UNETR aggressively down-sampling the 3D training images because of its heavy computational memory. SwinUNETR and UNETR decoders are equivalent to the 3D U-Net decoder, which utilizes transposed convolution for up-sampling and feature concatenation. Thus, any improvement in the performance of these models can be attributed to the increased capacity of the encoder to capture long-range spatial relationships.

### 2.2 Datasets:

This study applies in total 546 T2-weighted prostate-MRI volumes with whole-gland annotations derived from two independent sources:

Reader#1 Cohort (R#1, 342 volumes):

T2-weighted images for 342 patient samples were obtained from the public ProstateX challenge archive available on The Cancer Imaging Archive (TCIA) [22] and subsequently imported into our institutional research PACS (MIM Software®). Voxel spacing was resampled to $0.50 \times 0.50 \times 3.0$ mm³ for all subsequent processing. Two board-certified radiologists (12 years and 15 years of experience, respectively) in radiation oncology from the Moffitt Cancer Center contoured the whole prostate gland in three-dimensional planes with semi-auto contouring tools. The masks were kept in DICOM-RT format and transferred out for model training.

Reader#2 Cohort (R#2, 204 volumes):

A Separate expert annotated the second set of whole gland images for 204 ProstateX cases via TCIA [17]. These volumes are completely matched to the R#1 volume set and provide paired annotations for inter-reader comparison.

As a result, the utilized dataset includes 342 R#1 annotations and 204 R#2 annotations, a total of 546 label maps. At this stage both readers have annotated 204 standard training samples, and reader #1 has annotated a further 138 samples. The segmentation has been formulated as a binary classification, with prostate as foreground and the rest as background.

All TCIA data are completely de-identified; therefore, individual informed consent was waived. The study protocol underwent review and received approval from the University of South

Florida / Moffitt Cancer Center Institutional Review Board, and all procedures adhered to applicable human-subject research regulations.

### 2.3 Data pre-processing and augmentation:

All the applied T2-weighted images in this study were normalized and standardized by using the Monai deterministic approach. Where the MRI Images and ground truth masks were reoriented to align with the Right, Anterior, and Superior (RAS) directions, to facilitate the visualization, annotation, and model development for anatomical precision across all imaging planes [23]. Resampled to a standard 0.5 × 0.5 × 3.0 $mm^3$ grid, scaled to a 0–1 intensity range, and center-cropped or zero-padded to 128 × 128 × 32 voxels, pre-processing [24]. The normalized volumes were then fed through a stochastic augmentation block with random zooms (0.9–1.1), foreground-biased spatial crops (128 × 128 × 16) to address class imbalance, Gaussian blur/noise (p = 0.2 each), and 90° rotations along with flips (p = 0.5). Following this process, patches were resized back to 128 × 128 × 32 and converted to tensors [25]. This unified pipeline normalizes voxel geometry and integrates the data with realistic geometric and intensity variations leading to improved generalization, without introducing synthetically generated artifacts.

### 2.4 Loss functions and optimization:

Training the model reduces a composite loss that reconciles voxel-wise accuracy with regional overlap.

$$Loss = \frac{1}{2} L_{Dice}(w) + \frac{1}{2} L_{bce}(w)$$

$L_{Dice}$ represents the soft Dice loss, and, and $L_{bce}$ is the binary cross-entropy loss computed on each epoch to monitor the learning ability for each model. The combination of these two objectives, weighted by the inverse frequency of the voxels, is a common approach when dealing with the substantial foreground-to-background imbalance condition that may characterizing a specific samples of prostate gland MRI segmentation mainly the small gland size [26]. In practical application, $w$ are estimated to be the inverse of the voxel frequency of the training set in practice, where the trends of the foreground and background are around 5% and 95%, respectively, according to the approaches in [27, 28].

Hyperparameters were optimized utilizing Optuna version 3.3 [29], targeting three main hyperparameters (Learning rate $lr$, batch size, and feature size). The hyperparameter of the learning rate was optimized utilizing Optuna, which sampled from a categorical set consisting of six candidate values to reduce the computational cost. For each architecture, trained on six trial of discrete learning rate $(1 * 10^{-3}, 5 * 10^{-3}, 1 * 10^{-4}, 5 * 10^{-4}, 1 * 10^{-5}, 5 * 10^{-5})$, the batch size (1, 2, 4), and feature size (32, 48, 64, 96). The optimal hyperparameters for each dataset utilizing a specific Transformer model differ from those of other models. Specifically, the R#1 dataset has optimized parameters for the SwinUNETR of $(lr = 5 * 10^{-5}, batch\ size = 1, and\ feature\ size = 48)$ , In contrast, the best hyperparameters identified for the R#2 $(lr = 5 * 10^{-5}, batch\ size = 2, and\ feature\ size = 96)$. For the UNETR Transformer, the most effective training parameters for the R#1 dataset are $(lr = 1 * 10^{-3}, batch\ size = 2, and\ feature\ size = 48)$ while for the R#2 dataset, the parameters are $(lr = 5 * 10^{-5}, batch\ size = 1, and\ feature\ size = 96)$.

### 2.5 Evaluation protocol:

We evaluated the stability of the model under three different training strategies. (1) single reader training dataset with cross reader testing: for all the other experiments in this paper we have used the same amount of training dataset of 150 volume samples with varying number of training volume samples with 150, and test on remaining volume samples (R#1 → 192 test; R#2 → 54 test). The model was evaluated on its reader's test dataset as well as on the reader's adversarial testing dataset, demonstrating sensitivity to divergent annotation styles. (2) Mixed Training dataset with five-fold cross-validation technique: each dataset divided randomly into five folds training and testing datasets with (80% train 20% test per fold) for each, such that each and every case in the entire pool was present once and only once as a hold-out test set, resulting in an overall cross-site robustness score. (3) Gland size stress test: volumes were divided based on the average median gland volume for the datasets into 'large' and 'small' and one was fixed while the other was repeatedly down sampled to 85%, 70%, 55%, 40%, 25%, 25%, and 10% of its original size, with five-fold cross-validation and models trained repeatedly from scratch to evaluate robustness to extreme morphological imbalance. Overall performance was measured using the dice similarity coefficient (DSC), while differences between pairs of models were assessed with a two-tailed Wilcoxon signed-rank test (α = 0.05, Bonferroni-corrected). A schematic diagram of the process flow is given in Fig. 1.

Figure 1: Workflow to Train, infer and compare multiple prostate segmentation models using the Dice similarity coefficient and Statistics.

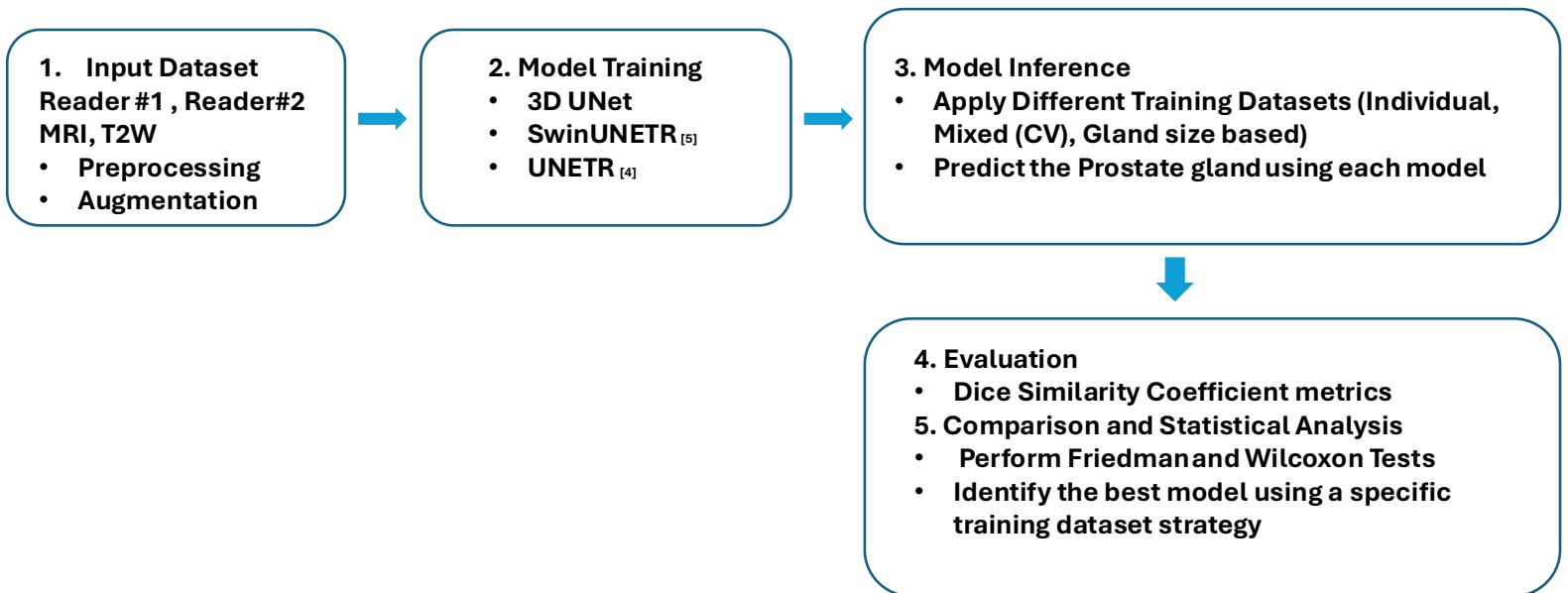

## 3. Result

In this section, we offer a comprehensive coverage of the segmentation performance of each model. We then examine three targeted analyses: cross-reader generalization, proportional reader mixing, and robustness to gland size. All reported numbers are the one on the test partitions unless otherwise indicated.

### 3.1 Overall accuracy:

The SwinUNETR and UNETR were trained using a single-cohort dataset independently; (R#1, n=150) and (R#2,n=150) were applied to each model, and we tested their performance on the testing dataset generated from both readers (Test R#1, n=192 & R#2,n=54). Where each model tested on each cohort testing dataset then evaluated on the cross-testing dataset, to study the model's performance on unseen testing dataset to evaluate the variability for the inter-reader dataset and study the ability of these models in adapting to these differences. According to our results, SwinUNETR outperformed other methods, achieving an average dice score of 0.82 and 0.817 when the model trained utilizing references from R#2 for R#1. For R#2 the average dice score of 0.87 on the same distribution of the training dataset and when the model tested when using references from other readers. However, for the trained UNETR, it fails to achieve the average Dice scores of 0.8 for R#1 and 0.8387 for R#2, and 0.7953 for R#1 and 0.842 for R#2, using the reference from the rest of the readers. The transformer models are also more stable in terms of prostate gland segmentation compared to the 3D UNet when subjected to cross-testing between the readers' testing data sets. (See Table 1)

Table 1: comparison of SwinUNETR, UNETR, and UNET Performance metrics R#1 and R#2 using single reader training dataset strategy.

| Model | | | | | | |
|---|---|---|---|---|---|---|
| **Training dataset** | SwinUNETR | | UNETR | | UNet | |
| | Testing Dataset | | | | | |
| | R#1, n=192 | R#2, n=54 | R#1, n=192 | R#2, n=54 | R#1, n=192 | R#2, n=54 |
| | Average Dice Coefficient | | | | | |
| **R#1, n=150** | 0.82015 | 0.817 | 0.8007 | 0.7953 | 0.825 | 0.7 |
| **R#2, n=150** | 0.8636 | 0.8695 | 0.842 | 0.8387 | 0.68 | 0.851 |
| **Lr** | $5 * 10^{-5}$ | $5 * 10^{-5}$ | 0.001 | $5 * 10^{-5}$ | $1 * 10^{-5}$ | |
| **Feature size** | 48 | 96 | 48 | 96 | | |
| **Batch size** | 1 | 2 | 2 | 1 | 22 | 5 |

### 3.2 Mixed Training dataset with five-fold cross-validation technique:

SwinUNETR and UNETR were trained and evaluated using the mixed training data set with five-fold cross-validation, which independently was conducted on datasets R#1 and R#2. the entire dataset of R#1, was firstly separated into five random folds of approximately 69 samples per fold and for R#2 entire dataset, was split into five folds of 40 samples per fold randomly. For the training part we concatenate the total training samples of (training R#1 &R#2, n= 436) which is the maximum training dataset in this study. The weighted average of five test folds DSC for each model are presented in Table 2. As shown in Table 2, the performance of SwinUNETR improved slightly for the R #1 and R #2 for the mixed training dataset in comparison with the single cohort training dataset results, the model yielded an average dice score of 0.8583 for the R#1 and 0.8678 for the R#2, demonstrating strong and consistent segmentation performance for both testing datasets. While UNETR performance improved significantly in comparison with the single dataset training experiment for both R#1 and R#2 testing dataset, achieving an average dice scores of 0.8513 and 0.8640 for R#1 and R#2 datasets, respectively, where the performance accuracy to segment prostate gland for SwinUNETR and UNETR is comparable.

The performance for the applied UNet model is also enhanced on this setup compared to a single cohort, but it performs slightly less accurately than the SwinUNETR and UNETR models, where the average dice score for R#1 and R#2 0.826 and 0.875, respectively. It reflects in a competitive but with lower performance over R#1, and the low but with a better segmentation performance for R#2 dataset. Performance of the models on the R#2 sample is presented in Figure 2. The presented comparison of the UNET (left), UNETR (center), and SwinUNETR (right) annotations. The ground truth label is shown in green, and predictions are shown in red. The mean dice scores of the models are 0.844, 0.876, and 0.896, respectively. On the contrary, for the patient R#1 sample, UNet has had some difficulty delineating the prostate gland, resulting in a mean dice score of 0.72, as illustrated in Figure 3. The UNETR model achieved a score of 0.76, and SwinUNETR surpassed this model with a mean Dice score of 0.833.

Table 2: The performance metrics comparison of the SwinUNETR, UNETR, and UNet models. Models were evaluated after being trained on the R#1 and R#2 cross-validation mix training dataset.

| **Model** | | | | | | |
|---|---|---|---|---|---|---|
| **CV-Training dataset** | SwinUNETR | | UNETR | | UNet | |
| **R#1, n= 274**<br>**R#2, n= 163** | Testing Dataset | | | | | |
| | R#1, n=68 | R#2, n=40 | R#1, n=68 | R#2, n=40 | R#1, n=68 | R#2, n=40 |
| | Average Dice Coefficient (μ, 95% CI) | | | | | |
| | 0.8583<br>[0.85 0.87] | 0.8678<br>[0.86 0.88] | 0.8517<br>[0.84 0.86] | 0.8640<br>[0.86 0.87] | 0.826<br>[0.81 0.84] | 0.875<br>[0.86 0.89] |

Figure 2: Representative slice of a patient's T2W image with delineation of Prostate gland provided by models UNet, UNETR, and SwinUNETR using mixed training dataset (R#1 & R#2) with fivefold cross validation shown on the R#2 sample T2-weighted MRI slice. Their scores are 0.844, 0.876, and 0.896, respectively. Ground truth is depicted in green, predictions in red.

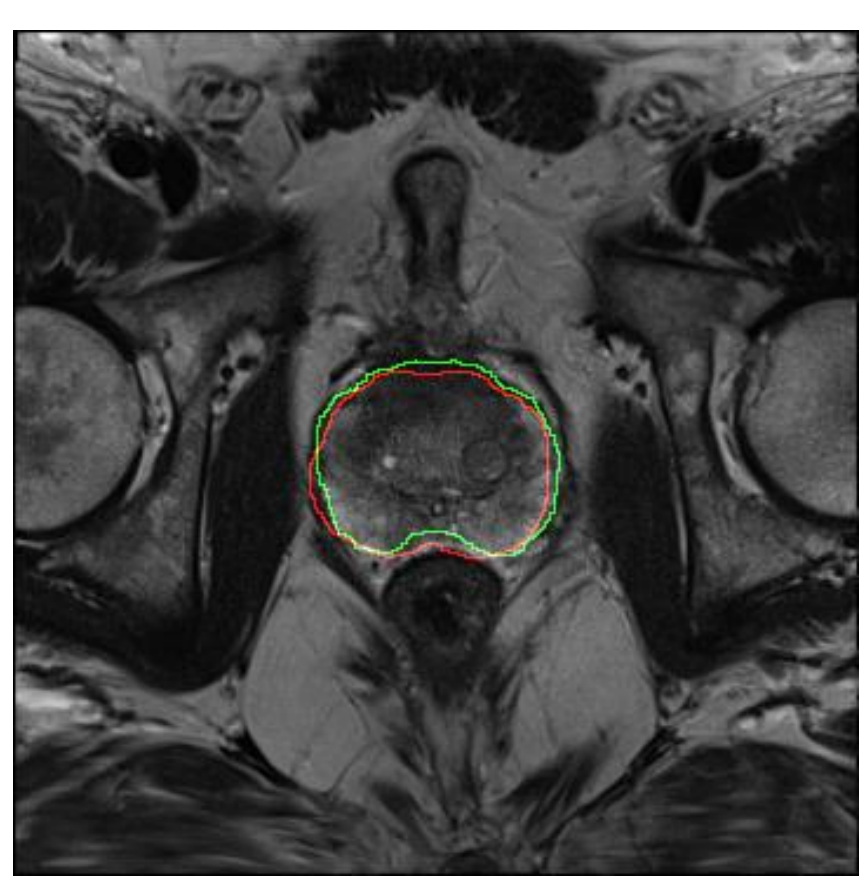

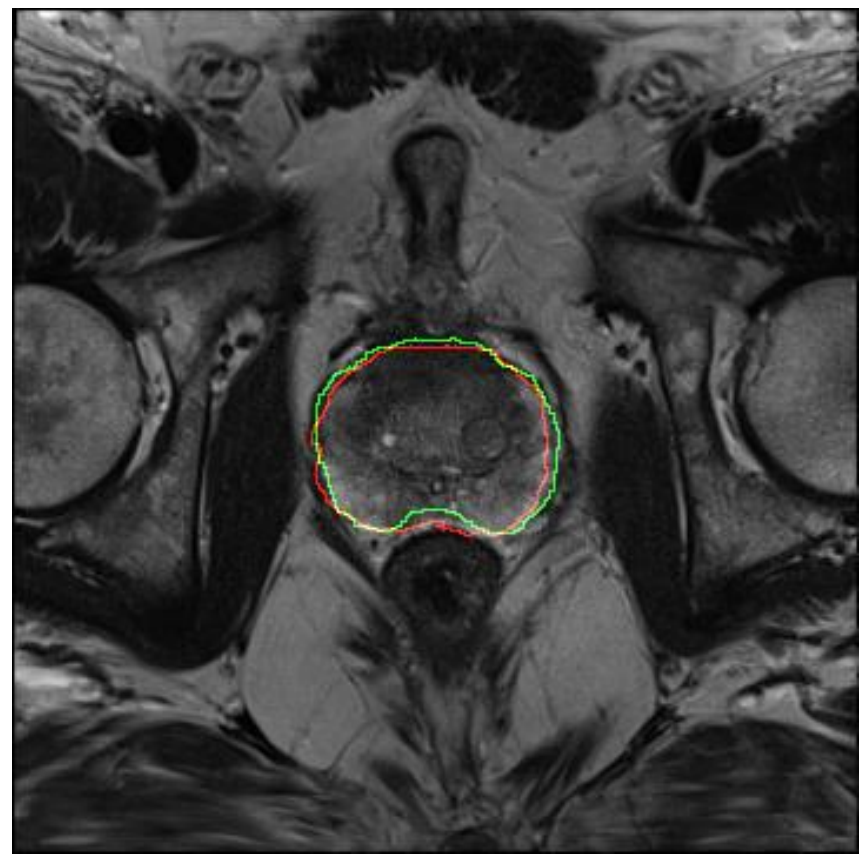

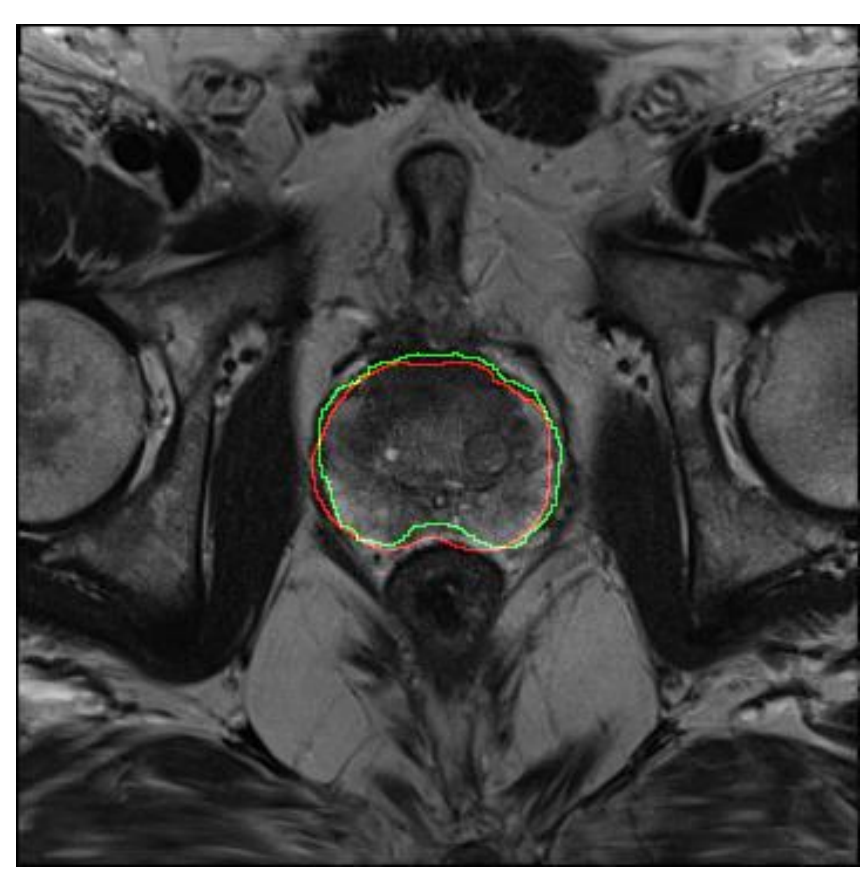

### 3.3 Robustness Based on a Gland Size-Based Dataset.

In this training strategy, we used the mixed training set method, combined with fivefold cross-validation, to train the model optimally for a fixed number of samples, as was done by fixing one dataset and decreasing the fraction of the second dataset. Performance comparison of Table 3A & 3B , each dataset, the large gland size-based subset dataset, when utilizing the entire dataset of the second dataset. For Table 3A we applied a proportional dataset of R#1 and fixed R#2 and evaluate the models performance for each case, in contrast we applied a proportional dataset of R#2 large gland based size dataset with the entire R#1 large gland-based size dataset, and qualitative overlay for three representative cases in figure 4, respectively.

Figure 3: Representative slice of a patient's T2W image with delineation of Prostate gland provided by models UNet, UNETR, and SwinUNETR using mixed training dataset (R#1 & R#2) with fivefold cross validation shown on the R#1 sample T2-weighted MRI slice. Their scores are 0.72, 0.76, and 0.83, respectively. Ground truth is depicted in green, predictions in red.

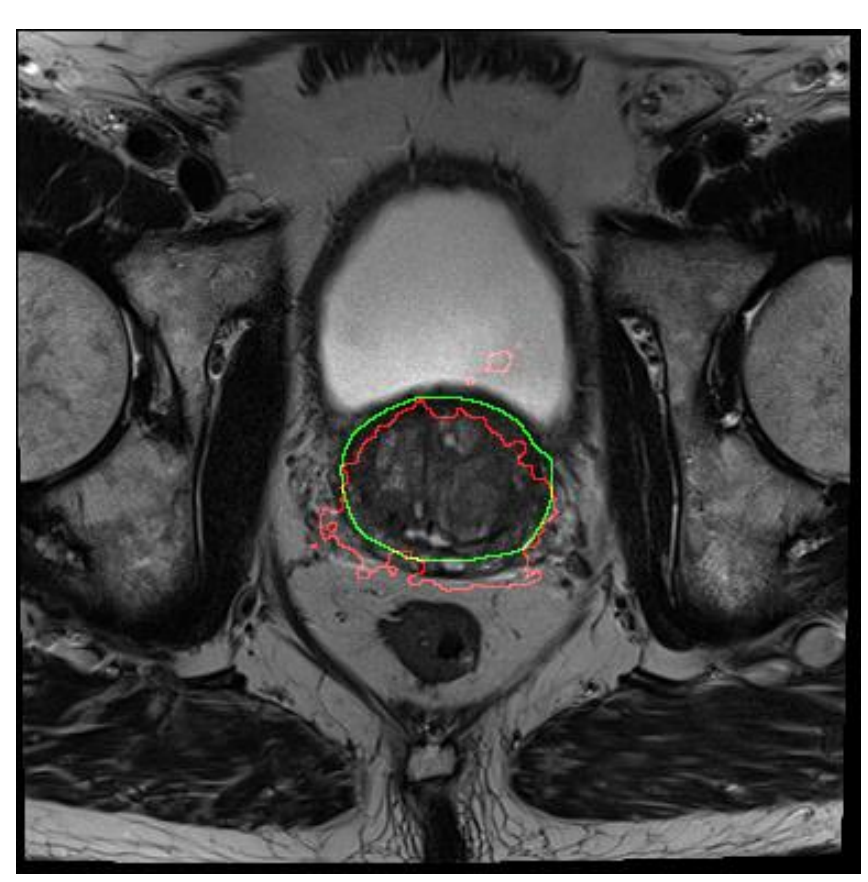

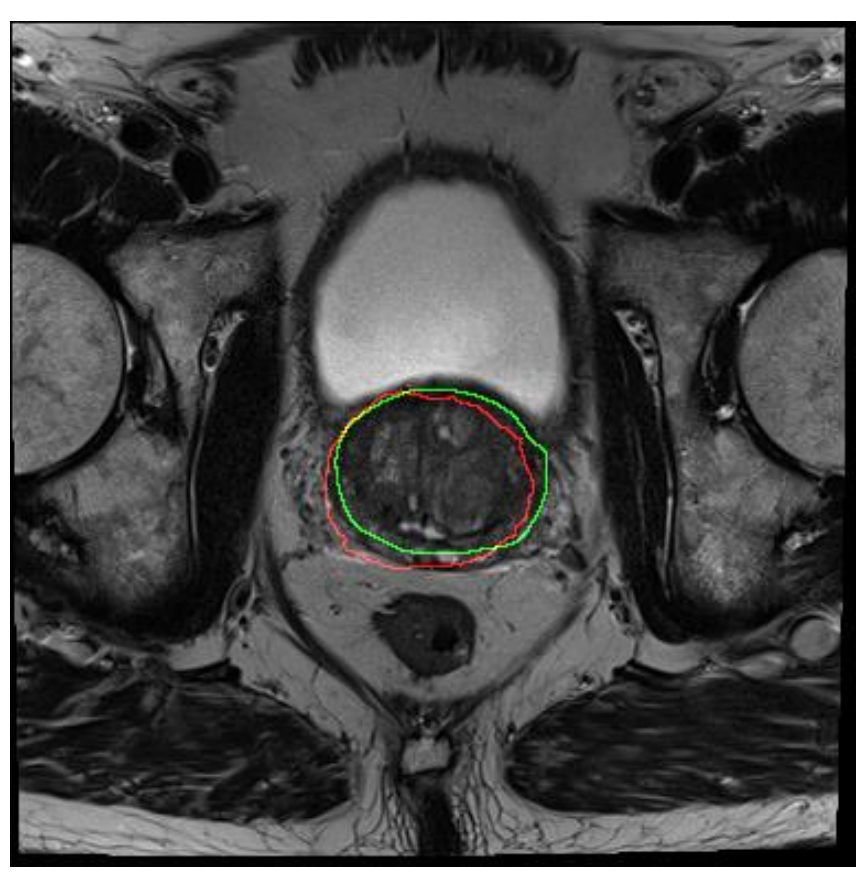

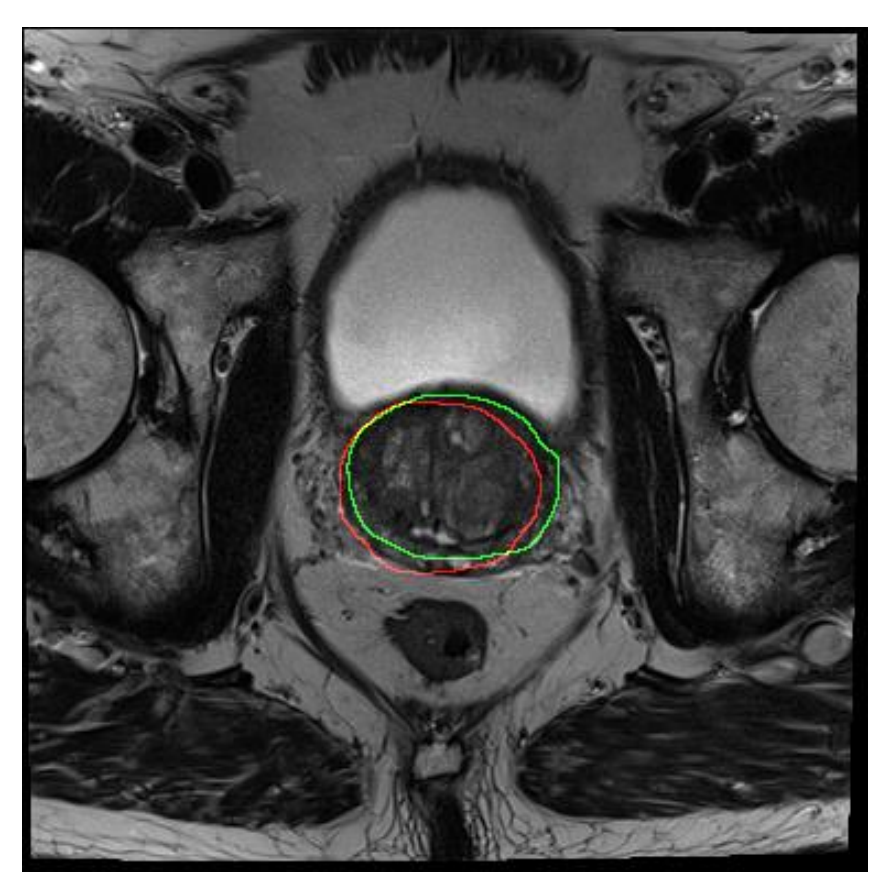

Table 3A shows the models' performance utilizing the entire R#2 large gland size dataset and a proportional R#1 for the large gland size dataset. We observe that the segmentation performance of SwinUNETR is the best, achieving superior average dice score of 0.879 to 0.902 on the large gland size R#1 testing dataset, while the UNet model achieved 0.846 as the maximum average dice score for R#1 dataset. For R#2, the average dice score between 0.889 and 0.894 on the R #2 test large gland size dataset, where the highest dice score for the UNet model 0.866. On the other hand, we have UNETR and then 3D UNet. For example, the maximum accuracy of the within-reader (R#1) is between 40% and 55%, proportional to the dataset, as SwinUNETR achieves an average dice score of 0.902 at 40% of R#1 applied. Furthermore, the average dice for R #2 tests remains strong, maintaining a value between 0.889 and 0.891 regardless of the amount of data from R #1 included, and the models generalize well to the other reader.

Table 3A: Model performance comparison on the large gland size-based proportional training dataset, where R#2 is fixed.

| **Model** | | | | | | |
|---|---|---|---|---|---|---|
| Training: R1 (%Proportion), R#2 fixed (n=87) | SwinUNETR | | UNETR | | UNet | |
| | Testing Dataset | | | | | |
| | R#1 | R#2 (n=22) | R#1 | R#2 (n=22) | R#1 | R#2 (n=22) |
| | Average Dice Coefficient (μ, 95% CI), 5-fold CV | | | | | |
| 100 | 0.871 [0.86 0.88] | 0.889 [0.88 0.9] | 0.8633 [0.85 0.87] | 0.886 [0.88 0.89] | 0.835 [0.82 0.85] | 0.859 [0.85 0.87] |
| 85 | 0.879 [0.87 0.89] | 0.891 [0.882 0.9] | 0.8723 [0.86 0.88] | 0.8893 [0.88 0.9] | 0.843 [0.83 0.85] | 0.862 [0.85 0.88] |
| 70 | 0.884 [0.87 0.89] | 0.891 [0.88 0.9] | 0.87611 [0.86 0.89] | 0.8865 [0.88 0.9] | 0.842 [0.82 0.86] | 0.866 [0.86 0.87] |
| 55 | 0.8878 [0.88 0.9] | 0.891 [0.88 0.9] | 0.8837 [0.87 0.89] | 0.8875 [0.88 0.9] | 0.84 [0.83 0.85] | 0.849 [0.83 0.86] |
| 40 | 0.902 [0.89 0.91] | 0.8942 [0.88 0.9] | 0.9013 [0.89 0.91] | 0.8849 [0.88 0.89] | 0.846 [0.82 0.87] | 0.849 [0.84 0.86] |
| 25 | 0.886 [0.87 0.9] | 0.8908 [0.88 0.9] | 0.8866 [0.88 0.9] | 0.8744 [0.86 0.88] | 0.838 [0.83 0.85] | 0.836 [0.82 0.86] |
| 10 | 0.8937 [0.88 0.91] | 0.8918 [0.88 0.89] | 0.889 [0.87, 0.9] | 0.867 [0.86 0.88] | 0.833 [0.8 0.87] | 0.838 [0.82 0.85] |

The variability for inter-reader datasets diminished significantly for almost all proportional cases for the transformer models, which is limited to 4 instances for the UNet model. In addition, the confidence interval percentiles are extremely narrow ±0.01, thus confirming that these performance increases and plateaus are statistically significant.

Observations for Table 3B SwinUNETR indicate that all R #2 training proportions yielded a mean dice score of between 0.887 and 0.896, UNETR with a mean Dice score of between 0.883 and 0.888, and UNet yielded a mean Dice score of between 0.815 and 0.857. Although SwinUNETR is able to achieve a slight drop of no more than 0.009 in Dice (from 0.896 to 0.887) using only 10% of R #2, it still proves to be remarkably robust. But UNETR and UNet observe relatively more drops.

We also investigated applying small-sized datasets by using the same way of training on a small-sized gland dataset with the models. Because Tables 4A & 4B are examples of more-detailed results by subset-testing dataset for R#1 and R#2, respectively.

Table 3B: Model performance comparison on the large gland size-based proportional training dataset, where R#1 is fixed.

| **Model** | | | | | | |
|---|---|---|---|---|---|---|
| Training: R2 (%proportion), R#1 fixed(n=134**)** | SwinUNETR | | UNETR | | UNet | |
| | Testing Dataset | | | | | |
| | R#1 (n=33) | R#2 | R#1 (n=33) | R#2 | R#1 (n=33) | R#2 |
| | Average Dice Coefficient (μ, 95% CI), 5-fold CV | | | | | |
| 100 | 0.871 [0.86 0.88] | 0.889 [0.88 0.9] | 0.8633 [0.85 0.87] | 0.886 [0.88 0.89] | 0.834 [0.82 0.85] | 0.866 [0.86 0.87] |
| 85 | 0.8689 [0.86 0.877] | 0.895 [0.88 0.9] | 0.8544 [0.84 0.87] | 0.8868 [0.87 0.91] | 0.838 [0.83 0.84] | 0.872 [0.86 0.89] |
| 70 | 0.868 [0.86 0.88] | 0.8948 [0.89 0.9] | 0.8531 [0.84, 0.86] | 0.8883 [0.88 0.9] | 0.823 [0.81 0.83] | 0.866 [0.85 0.88] |
| 55 | 0.865 [0.86 0.87] | 0.8879 [0.88 0.9] | 0.8515 [0.84 0.86] | 0.8845 [0.88 0.89] | 0.832 [0.82 0.84] | 0.870 [0.86 0.88] |
| 40 | 0.8622 [0.85 0.87] | 0.8866 [0.88 0.9] | 0.846 [0.84 0.86] | 0.8795 [0.87 0.89] | 0.825 [0.82 0.83] | 0.869 [0.84 0.89] |
| 25 | 0.8636 [0.86 0.87] | 0.8824 [0.87 0.9] | 0.8443 [0.83 0.85] | 0.8726 [0.86 0.89] | 0.827 [0.82 0.84] | 0.86 [0.84 0.88] |
| 10 | 0.8621 [0.85 0.87] | 0.8961 [0.88 0.91] | 0.8422 [0.83 0.85] | 0.8838 [0.87 0.9] | 0.815 [0.8 0.83] | 0.857 [0.78 0.93] |

For the large gland size datasets, we observed that both Transformer-based models maintain their performance stability and ability to segment the prostate gland, regardless of the number of proportional samples from the applied dataset, which is not the case for the UNet model.

The analysis in Table 4A shows that we used this small gland size sub dataset for our analysis, where we reduced the R#1 datasets from 100% to 10% and used the whole R#2 small gland size dataset. R#1 dices score for all models, the same approach reduces the R#1 dices score. For all models, the best performance for SwinUNETR was obtained with 40% from R#1 small gland size and 100% from R#2 small gland with an average dice score of 0.8494, and the average dice score was 0.841 for the UNETR, and 0.8 for the UNet model. The SwinUNETR achieves a consistent average dice score (0.85±0.003) across all R#1 proportion scenarios to R#2. While on the other hand, at the extreme 10% of the R#1 small gland size data set, the UNETR model also notes a decrease from 0.85 to 0.814 with the same change of -0.036. In addition, the UNet framework is the most sensitive to the imbalance, as it shows a decrease from 0.831 to 0.787 and thus a change of about -0.044. Our transformer backbones used in this work also show much better generalization than the conventional UNet under the extreme class/domain imbalance situations.

Table 4A: Model performance comparison on the small gland size-based proportional training dataset, where R#2 is fixed.

| **Model** | | | | | | |
|---|---|---|---|---|---|---|
| Training: R1 (%Proportion), R#2 fixed (n=76) | SwinUNETR | | UNETR | | UNet | |
| | Testing Dataset | | | | | |
| | R#1 | R#2 (n=19) | R#1 | R#2 (n=19) | R#1 | R#2 (n=19) |
| | Average Dice Coefficient (μ, 95% CI), 5-fold CV | | | | | |
| 100 | 0.8477 [0.84 0.86] | 0.8537 [0.84 0.87] | 0.836 [0.83 0.85] | 0.85 [0.84 0.86] | 0.809 [0.8 0.82] | 0.831 [0.82 0.85] |
| 85 | 0.8482 [0.84 0.86] | 0.8475 [0.83 0.86] | 0.8383 [0.83 0.85] | 0.8482 [0.83 0.86] | 0.803 [0.78 0.83] | 0.829 [0.81 0.85] |
| 70 | 0.8484 [0.84 0.86] | 0.8542 [0.84 0.87] | 0.8411 [0.83 0.85] | 0.8484 [0.83 0.86] | 0.798 [0.77 0.82] | 0.813 [0.79 0.84] |
| 55 | 0.8472 [0.83 0.86] | 0.8508 [0.84 0.86] | 0.8426 [0.83 0.86] | 0.8472 [0.82 0.85] | 0.798 [0.78 0.82] | 0.817 [0.81 0.83] |
| 40 | 0.8494 [0.83, 0.86] | 0.8506 [0.84 0.86] | 0.841 [0.83 0.86] | 0.8494 [0.81 0.84] | 0.8 [0.79 0.81] | 0.809 [0.79 0.83] |
| 25 | 0.843 [0.82 0.86] | 0.85099 [0.84 0.86] | 0.8379 [0.82 0.86] | 0.843 [0.81 0.84] | 0.805 [0.79 0.82] | 0.798 [0.78 0.81] |
| 10 | 0.814 [0.78 0.85] | 0.8499 [0.84 0.86] | 0.8174 [0.78 0.85] | 0.814 [0.81 0.83] | 0.754 [0.74 0.77] | 0.787 [0.75 0.82] |

In contrast, the results in Table 4B, SwinUNETR, show significantly higher robustness to the limited access of R#2. For the small gland size testing dataset in which the ratio of R#2 is 10%, the average dice score for R#1 and R#2 is 0.846 and 0.847 respectively indicating the robustness of the model. In comparison, UNETR suffers the steepest drop in performance as R#2 decreases, hitting peak model performance with the full small gland size mixed training dataset, for which it obtains an average Dice score of 0.836 for R#1 and 0.85 for R#2. The UNet model shows mild damage here, as it performs worse than SwinUNETR but better than UNETR relatively when faced with a large imbalance.

Table 4B: Model performance comparison on the small gland size-based proportional training dataset, where R#1 is fixed.

| **Model** | | | | | | |
|---|---|---|---|---|---|---|
| Training: R2 (%proportion), R#1 fixed (n=140) | SwinUNETR | | UNETR | | UNet | |
| | Testing Dataset | | | | | |
| | R#1 (n=35) | R#2 | R#1 (n=35) | R#2 | R#1 (n=35) | R#2 |
| | Average Dice Coefficient (μ, 95% CI), 5-fold CV | | | | | |
| 100 | 0.8457 [0.84 0.86] | 0.8472 [0.84 0.87] | 0.836 [0.83 0.85] | 0.85 [0.84 0.86] | 0.808 [0.79 0.83] | 0.826 [0.81 0.84] |
| 85 | 0.8435 [0.84 0.86] | 0.8493 [0.83 0.86] | 0.7647 [0.75 0.78] | 0.7842 [0.77 0.8] | 0.803 [0.79 0.82] | 0.817 [0.79 0.84] |
| 70 | 0.841 [0.83 0.85] | 0.8478 [0.83 0.87] | 0.827 [0.82 0.84] | 0.8524 [0.84 0.87] | 0.804 [0.79 0.81] | 0.837 [0.82 0.85] |
| 55 | 0.8365 [0.83 0.85] | 0.8438 [0.83 0.87] | 0.8148 [0.81 0.83] | 0.8438 [0.83 0.86] | 0.792 [0.78 0.81] | 0.824 [0.81 0.84] |
| 40 | 0.8333 [0.83 0.85] | 0.8161 [0.82 0.86] | 0.8099 [0.8 0.83] | 0.8146 [0.82 0.87] | 0.795 [0.79 0.81] | 0.841 [0.82 0.86] |
| 25 | 0.835 [0.82 0.84] | 0.8153 [0.79 0.84] | 0.81 [0.8 0.82] | 0.8059 [0.79 0.84] | 0.802 [0.79 0.81] | 0.824 [0.8 0.84] |
| 10 | 0.8457 [0.82 0.85] | 0.8472 [0.77 0.86] | 0.7674 [0.8 0.82] | 0.7841 [0.74 0.87] | 0.794 [0.78 0.81] | 0.808 [0.77 0.85] |

Moreover, we present a comprehensive comparison of prostate gland segmentation performed by the transformer models to understand the performance differences between these models, using a single patient MRI sample. At the top, the efficacy of SwinUNETR in delineating the prostate gland is highlighted, whereas the efficacy of UNETR is depicted at the bottom. Figure 4 represents the utilized training strategy of the mixed training of the small gland size dataset of R#1 and R#2, along with the cross-validation based on a 40% sample of R#2 and the entire R#2 small gland dataset (R#1, n=56; R#2, n=76).

In the implemented training strategy shown in Figure 4, it appears that for the same patient sample, we present three sequential slices to present the model annotation in comparison with the ground truth annotation for R#2 sample, SwinUNETR presents smoother and more continuous boundaries for the prostate gland than the UNETR performance. Mix-proportional small-gland size training dataset with cross-validation: UNETR shows both over-segmentation and under-segmentation artifacts this as vision evidence for the case of UNETR sensitivity to class imbalance for the small gland size samples in which the UNETR performance reduces significantly . Proportional mixed training in the context of the small-sized gland training strategy provides a good overlap with contours defined by experts when combined with SwinUNETR. This means it is the best part for achieving strong prostate segmentation for all of the training strategies described.

Figure 4: Representative slice of a patient's T2W image with delineation of Prostate gland provided by (A) SwinUNETR (B) UNETR with mixed training dataset with CV.

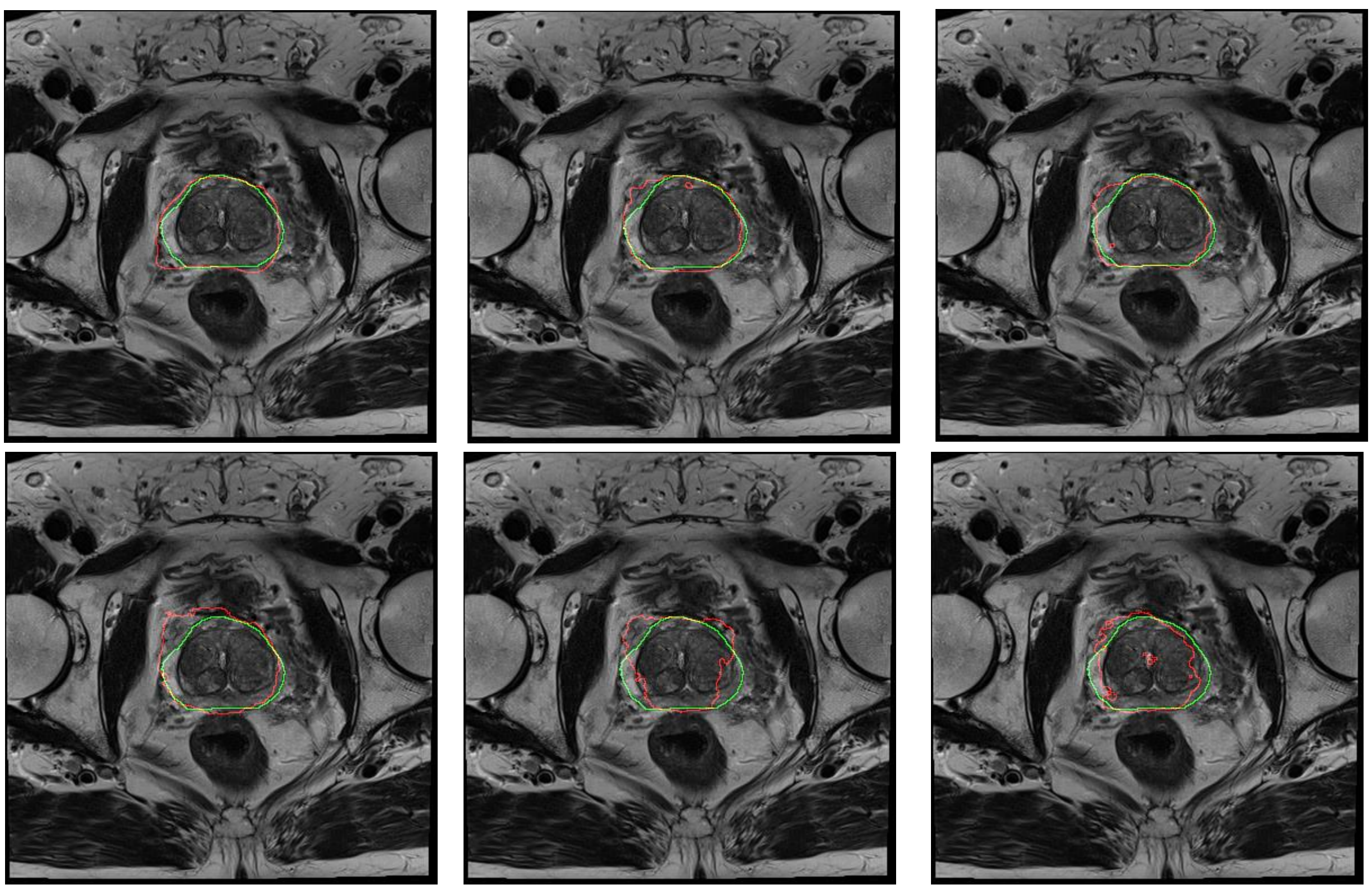

Figure 5: Statistical analysis of Prostate segmentation results for each of the three models (UNet, SwinUNETR, and UNETR) on R#1 and R#2 single-reader datasets.

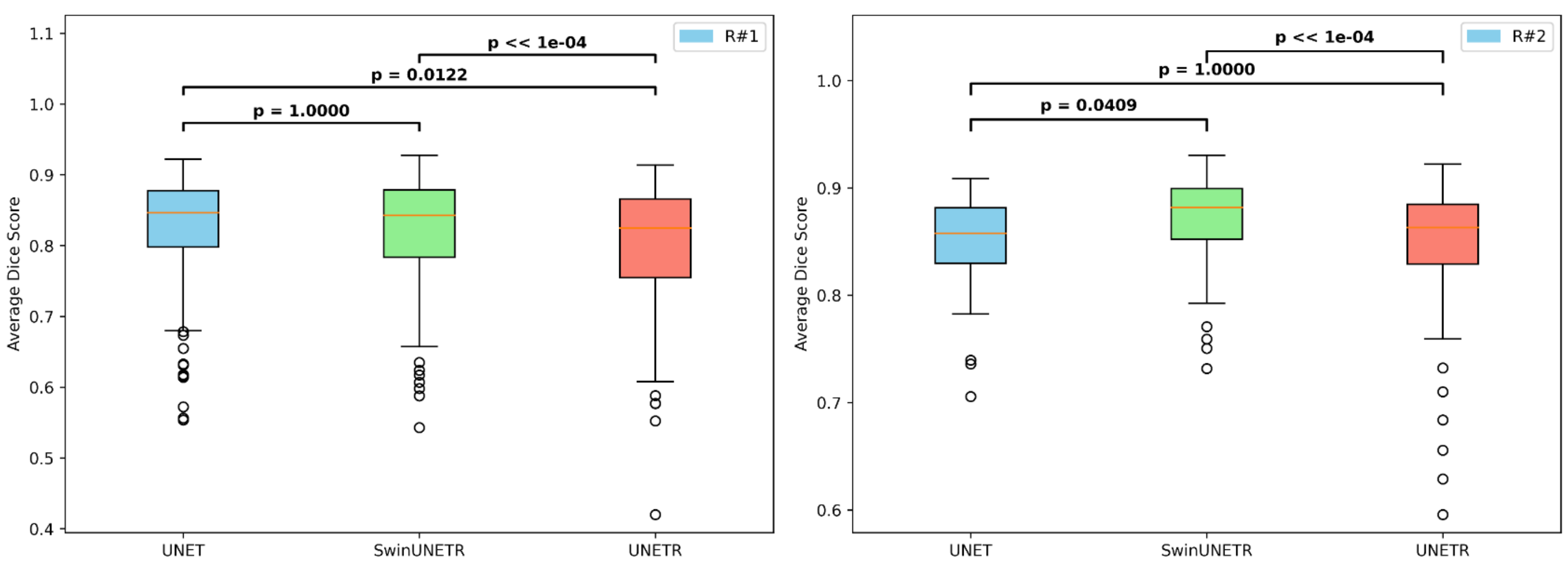

In order to conduct a more rigorous and formal analysis of the influence of model selection on the applied datasets R#1 and R#2, in Figure 5, we employed the Friedman test to compare training strategies and to test for the presence of significant differences in model performances. Within the single training strategy for R#1, we found that the performance of SwinUNETR compared to UNETR is significantly different, with a p-value less than 0.0001. Furthermore, the performance difference between 3D U-Net and UNETR is also significant; however, the performance difference between SwinUNETR and 3D U-Net is not significant. Regarding R#2, the results are consistent with those for R#1, indicating that the performance difference between SwinUNETR and UNETR is statistically significant, whereas there is no significant difference when comparing the performance of 3D U-Net and UNETR, see Figure 5.

In the training strategy based on the mixed training dataset with cross validation, the Friedman test showed a p-value <0.0001 for the reader (R#1) across the three models. Therefore, it shows that the three models' performance distributions are certainly not the same. With R#2, the Friedman test then yielded a p-value greater than the threshold value 0.05, then it is not significant to apply any of these models on R#2 dataset within the same training strategy and showing that the SwinUNETR and UNETR perform equally on R#2 testing datasets.

Figure 6: Statistical analysis of Prostate segmentation results for each of the three models (UNet, SwinUNETR, and UNETR) on R#1 and R#2. Mixed training dataset with CV .

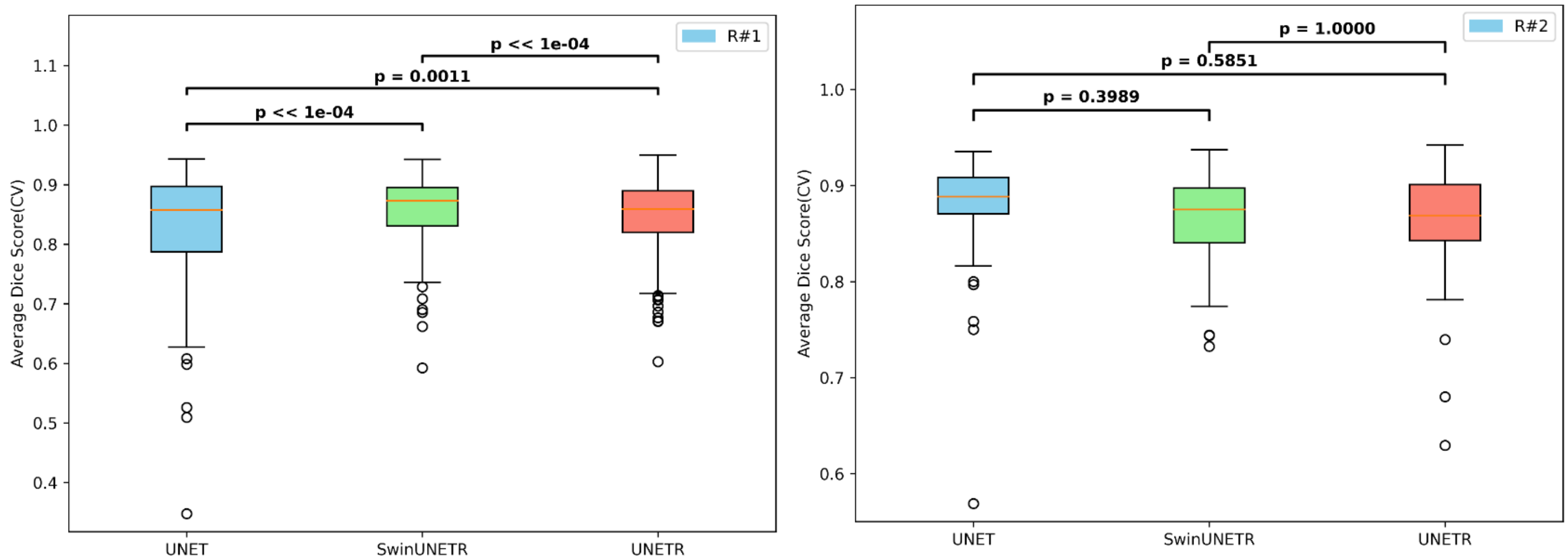

Additionally, we applied the same comparison using the mixed training dataset with a cross-validation technique, where SwinUNETR leads the model in boosting performance on R#1, which has a significant difference compared to 3D U-Net and UNETR. Still, when we test the performance for R#2, we found that these results are not valid. In this case, the characteristic of the dataset plays a crucial role in affecting the model's performance when it is exposed to the R#2 dataset mixed with the R#1 dataset, see Figure 6.

Another investigation on mixed training dataset with five-fold cross-validation on R#1 and R#2 is used for the model selection, in which SwinUNETR and UNETR are used. Boxplot analysis of the Dice similarity coefficients showed that segmentation accuracy was significantly improved ($p \ll 1e\text{-}3$), with SwinUNETR providing the best agreement between predicted and the true prostate gland outlines on the R#1 dataset. For R#2 data set, this is not the case because the p-value is equal to 0.379, which means similar performance. These results allow us to infer a sense of stability and robustness of the SwinUNETR behavior over the clinical dataset on which we apply it. Other datasets might exhibit different behavior showing that the properties of the dataset influenced the behavior of the model, see Figure 7.

Figure 7: Comparison of segmentation performance of SwinUNETR and UNETR with mixed training dataset with CV on R#1 and R#2 testing datasets.

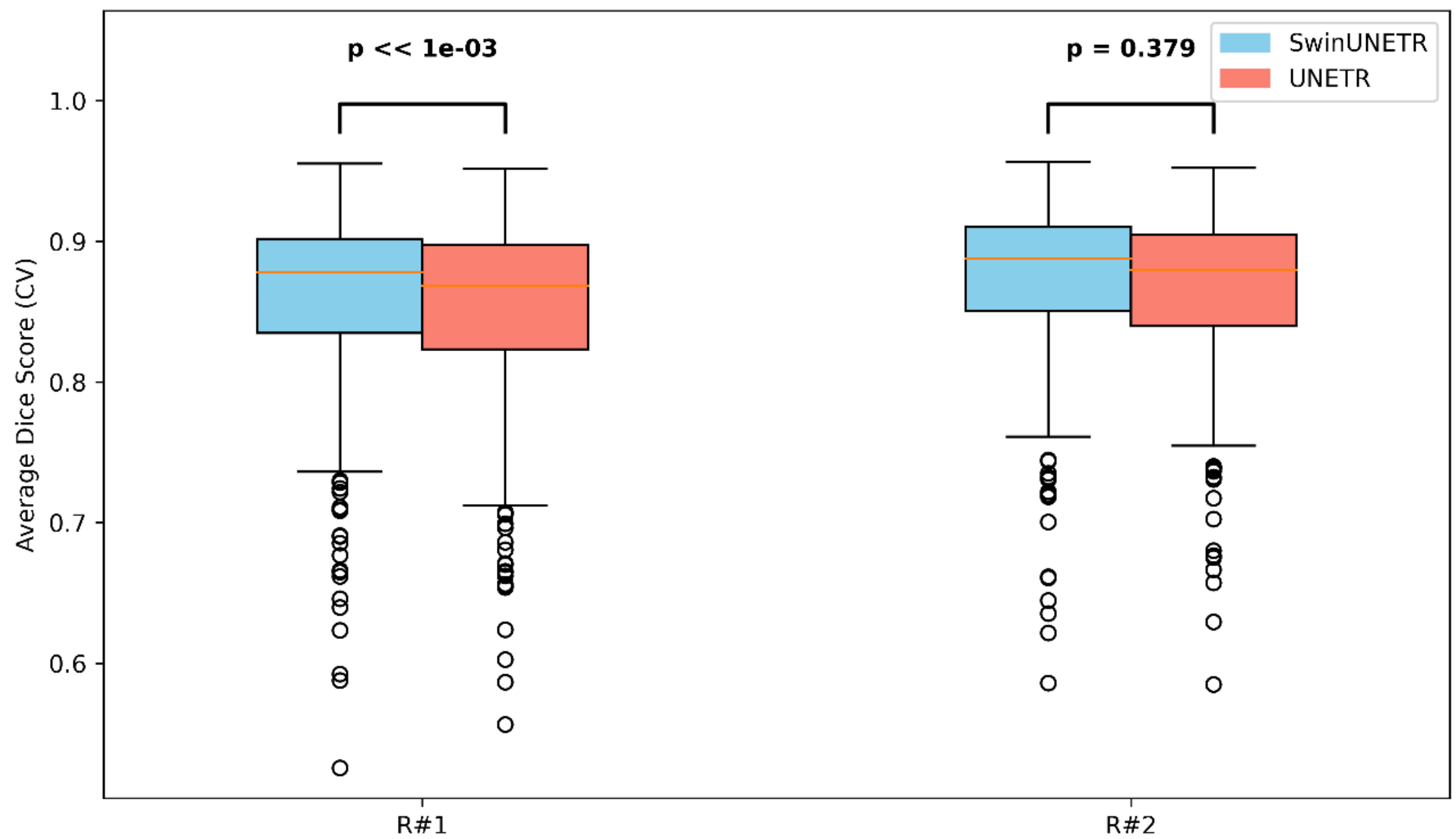

It can be derived from the prior results that the transformer-based SwinUNETR and UNETR models' performance are steady and consistent in segmenting prostate gland on the dataset where it is applied. These models exhibit a similar degree of strength, especially for the R#1 dataset. Yet, the UNET model remains competitive amongst the R#2 dataset. While SwinUNETR consistently demonstrates excellent accuracy across a wide span of values of the number of training samples, the performances of UNETR and UNet depend on the content of the training data (a mixture of classes).

## 4. Discussion:

This study utilizes three deep learning models in order to 1) systematically compare their performance at delineating the prostate gland and 2) their efficacy in minimizing variability across inter-reader datasets. We evaluated our results with three separate experiments on MRI (T2-weighted): (1) A single reader training cohort, (2) Mixed reader training with cross-validation and (3) a proportional of the dataset based on the size of the prostate gland. Using SwinUNETR, UNETR, and UNet models, and a series of applied training strategies we systematically investigate model robustness, generalizability, and sensitivity across varying reader annotations, and prostate gland characteristics.

Nowadays, the transformer-based models with attention mechanism utilized in top their successful in natural language detection [30] , and applied to image segmentation with high accuracy through implementing the vision transformers (ViT) [31], there is a significant interest in the application of transformer based encoders specifically designed for medical image segmentation task [32]. The inductive bias [33, 34] of the transformers is weaker than that of the convolutional neural network models since the transformers employ self-attention mechanisms. An example of the hybrid segmentation transformer model is SwinUNETR, which used the Swin Transformer encoder . The approach of this model is to capture global context first using an attention mechanism called shifted window attention and then reconstructs a complex spatial feature from convolutional skip connections. [35, 36] This makes this approach less sensitive to images of different scales, compared to traditional convolution-based techniques, which is an added benefit of this methodology. [37] As such, it has shown remarkable resilience for accurate medical imaging segmentation of complex anatomical structures in the 3D medical images.

The UNETR Transformer model integrates a Transformer-based encoder module with a convolutional decoder to exploit from the proprieties of the transformers and bias ability of the convolutions for 3D medical image segmentation. The encoder chunks the 3D input volumes into unique, non-overlapping 3D patches, treating each non-overlapping patch as a self-attention mechanism which learns global context and long-range dependencies [38]. In addition, the UNETR applied on complex organ segmentation and improved the accuracy by at least with 1 percent for the applied organs[39]. The convolutional decoder then up-samples these features while using skip connections to blend some of the spatial information from the encoder. This architecture achieves segmentation performance by jointly modeling global context and reconstructing local features to provide better contextual information for the segmentation task and apply a segmentation-based architecture in medical imaging task.

These models have been widely used for 3D medical segmentation, especially for the prostate volume. The high variability of the prostate gland in terms of its shape and texture, coupled with the intertumoral heterogeneity of prostate cancer, makes the segmentation of the prostate gland one of the great challenges in the development of automated models [40]. We trained SwinUNETR and UNETR Transformer models employing inter-reader datasets to evaluate their performance in comparison with our prior results utilizing 3D UNet. Using inter-reader datasets

SwinUNETR and UNETR Transformer models were trained to assess their performance against results previously achieved with 3D UNet, cited in [1] , using a much larger cohort (n=345) that enabled enhanced training and evaluation of models via Optuna, an open-source framework for hyperparameter optimization, which automates the process of determining optimal hyperparameters to enhance model performance. This approach uses sophisticated search algorithms and early stopping criteria, saving capital [41]. We implemented specific training strategies to address the limitations identified in our previous study.

Model [42] consists of a 2D organ volume localization network, followed by a 3D segmentation network to volumetrically segment the prostate. The Dice score achieved for the prostate gland is approximately 0.9 with respect to 136 patients, contend the authors. As such, the authors state that the average Dice score that nnU-Net gets is 0.855 (70 to 0.95) in [26] utilizing the Promise12 public dataset [43]. However, locally, the segmentation of the prostate gland and lesion performed better with the Transformer Encoder Decoder [44] , over four convolution-based deep learning neural models.

In summary, we trained and evaluated two transformer-based three-dimensional segmentation models, UNETR and SwinUNETR, for the human prostate gland delineation from T2-weighted MRI scans. The comparison was comprised of careful benchmarking with a standard 3D model, specifically with a 3D UNet. An inter-reader dataset was used, and two experts from different institutions annotated each volumetric image independently. In addition, we compared performance metrics across a variety of training configurations using a single-reader setting, a mixed-reader setting, and a proportional-mix configuration. For all tested configurations the transformer-based models outperformed the 3D UNet yielding higher Dice scores and narrower confidence intervals. Of particular note is the most compact model, SwinUNETR, which provided the most consistently stable results in the face of differing annotation methods. In addition, qualitative evaluation also found that the capsular boundaries delineated by the attention-based models were smoother with fewer false positives, reinforcing their improved ability to capture the prostate global context.

Limitation: While inter-reader variability is addressed with readers from other institutions, this study remains limited in the T2-weighted sequences that were performed on a smaller set of scanners and vendors. Multiple annotator datasets will theoretically lead to better generalization for these models, and applying another hybrid transformer model should contribute to model stabilization and agreement.

## 5. Conclusion:

In this study, we find that transformer-based encoders largely outperform the 3D UNet baseline on an analysis of multi-reader prostate magnetic resonance imaging (MRI). This gives rise to 5 percentage points increase of whole-gland Dice score by UNETR and SwinUNETR. Importantly, despite being trained on an extremely imbalanced dataset, SwinUNETR retains its ability to bridge inter-reader performance variability, and it is also the most robust to the effect of annotation variability. These results indicate that global self-attention can lessen expert noise and pave the way for reader-agnostic segmentation models that are generally applied between radiologists or institutions. In addition, due to an excellent balance of accuracy and computational efficiency, SwinUNETR generates fully-automated prostate contours on standard graphics processing units (GPUs) in near–real-time, making it an attractive choice for use in the clinical routine.